\documentclass[10pt,twocolumn,article]{IEEEtran}
\usepackage{times,comment}
\usepackage{amsbsy}
\usepackage{latexsym} 
\usepackage{amssymb}
\usepackage{enumitem}
\usepackage{mathtools,xparse,nicefrac}
\usepackage{amsmath,amsfonts,graphicx,epsfig,amsthm,mathtools,diffcoeff}
\DeclarePairedDelimiter{\norm}{\lVert}{\rVert} \usepackage{times,latexsym,bm,color}
\usepackage{algpseudocode, algorithm}
\usepackage{setspace}

\begin{document}
\title{\huge {Study of Multiuser Multiple-Antenna Wireless Communications Systems Based on Super-Resolution Arrays }}
\author{Silvio F. B. Pinto and Rodrigo C. de Lamare \vspace{-2.00em} 
\thanks{The authors are with the Center for Telecommunications Studies (CETUC), Pontifical Catholic University of Rio de Janeiro, RJ, Brazil. R. C. de Lamare is also with the Department of Electronics, University of York, UK. Emails: silviof@cetuc.puc-rio.br, delamare@cetuc.puc-rio.br}}
\maketitle 
\begin{abstract}
This work studies multiple-antenna wireless communication systems based on super-resolution arrays (SRAs). We consider the uplink of a multiple-antenna system in which users communicate with a multiple-antenna base station equipped with SRAs. In particular, we develop linear minimum mean-square error (MMSE) receive filters along with linear and successive interference cancellation receivers for processing signals with the difference co-array originating from the SRAs. We then derive analytical expressions to assess the achievable sum-rates associated with the proposed multiple-antenna systems with SRAs. Simulations show that the proposed multiple-antenna systems with SRAs outperform existing systems with standard arrays that have a larger number of antenna elements.
\end{abstract}

\vspace{-1.00em}
\section{Introduction}
In the last decade, multiple-antenna technology \cite{mmimo,overview,wence} has attracted the interest of researchers dedicated to the investigation of fifth generation (5G) mobile networks due to its substantial improvement in the capacity of wireless systems and networks. In the specific case of the multiple access channel, multiple-antenna systems employ a number of physical antennas at the base station (BS),  which is often much larger than that of the total number of users. However, the higher spectral efficiency and resolution resulting from this excess of antennas comes at the expense of physical space, coupling effects \cite{Gupta,Wu} among antenna elements and energy consumption \cite{Pinto1}. Among some innovative solutions proposed to overcome these drawbacks are coarse quantization \cite{Pinto2}, which  provides energy savings. Another is the use of sparse or super-resolution arrays (SRAs) \cite{Pal1,Liu1,Liu2,Wang_coprime,wesley,imisc,emisc} in the BS whose large degrees of freedom can be obtained by the virtual expansion of the number of  physical sensors. 

The essential feature that allows one to obtain more sensors than the physical ones leads to more compact antenna arrays, reduced electromagnetic coupling  \cite{Liu1,Liu2} effects and significant energy savings.
It is known that  among the  known types of sparse arrays such as nested, co-prime, minimum redundancy  (MRAs) and minimum hole  (MHAs) only the first and the second  provide simple closed-form expressions for the array geometry and their sensor locations must be found from tabulated entries.  Despite  the co-prime arrays yield virtual ULAs consisting of smaller number of  sensors than nested arrays and the MRA,  they  result in useful virtual ULAs. Therefore, it is desirable to take advantage of the properties of the mathematical properties common to both in order to formulate a unified approach for exploiting the benefits of their unequal spacing and virtual increase of sensors in multiple-antenna systems to increase achievable sum rates and accuracy. 
Previous works \cite{Feng} studied nested arrays technique in a massive multi-input multi-output 
(MIMO) heterogeneous network and the problem of joint user association and interference nulling scheduling to maximize the sum rate of users of small and macro-cells,  investigated its channel estimation \cite{Zhu} for massive MIMO in 2-D and extended to 3-D  by \cite{Yuan}. However, neither of them focus on the possibility of unifying the processing of sparse arrays to  apply them to MIMO systems. 

In this work, we investigate the uplink of multiuser MIMO systems based on SRAs and develop receivers for processing signals with the difference co-array from the SRAs. In particular, we consider co-prime and two level nested arrays, in a single procedure that can be applied to multiuser MIMO systems from the point of view of the properties of the similar virtual ULAs obtained by preprocessing each one. We adopt a {tailored geometry-based stochastic model \cite{Clerckx,Hawej,Hawej_2,Hawej_3,Zheng} that preserves the structure of the steering vectors characterized by non-uniform delays, which is a sine qua non condition \cite{Pal_1,Pal_2} to obtain increased number of virtual sensors with these methods. This is provided by assuming no scattering for the propagation inside the single-cell and  also that each path is associated to its user and its respective steering vector.} We then derive linear minimum mean-square error (MMSE) receivers for processing signals with the difference co-array with super resolution. The  performance appraisal of the proposed super resolution multiple-antenna  processing through longstanding metrics shows that substantial gains can be achieved in terms of savings in the number of antennas and its resulting reduction of energy  consumption, increased achievable sum-rates and improved bit error rate (BER), which encourages further research in this field.     

This paper is structured as follows. Section II describes the system model and background for understanding the proposed
super-resolution processing. Section III presents the proposed
super-resolution multiple-antenna processing. Section IV formulates the proposed
MMSE receiver, including  its resulting Successive Interference Cancellation form, for Two-level Nested Arrays (TLNA) and Co-prime arrays (CPA), whereas Section V analyzes aspects of the sum-rate performance of the
proposed super-resolution multiple-antenna processing 
Section VI presents and discusses numerical results whereas the conclusions are drawn in Section VII.
%
%
\vspace{-1.0em}
\section{Definitions and Multiple-Antenna System Model}
\vspace{-0.6em}
In this section, we review some basic on sparse arrays to help the understanding of the subsequent material and then introduce the multiple-antenna system model. We assume that the multiple-antenna system model under consideration employs a sparse antenna array at the base station (BS). 
\subsection{Definitions}
\vspace{-0.5em}
{Definition 2.1.  The difference coarray set \cite{Pal_1}represented by $\mathcal{D}$ is a set associated with the sensors' positions  $\mathcal{S}$ through}
\begin{equation}
 \mathcal{D} \triangleq \left\{\mathit{n}_{1}- \mathit{n}_{2} \mid \left(\mathit{n}_{1}, \mathit{n}_{2} \right) \in  \mathcal{S}^{2}\right\}
\end{equation}
Definition 2.2.  The number of Degrees of
Freedom \cite{Pal_1}, denoted by $\mathit{DoF}$, of a geometry specified by $\mathcal{S}$ is the cardinality of its difference coarray set, as follows:
\begin{equation}
\mathit{DoF}\triangleq \mathcal{D}
\end{equation}
Definition 2.3. A two-level nested array \cite{Pal_1}is a  sparse array consisting of the union of the sensors of two ULAs. The inner $\mathcal{S}_{in}$ presents $\mathit{M}_{1}$ sensors and the external $\mathcal{S}_{ou}$  possesses $\mathit{M}_{2}$ sensors. Their locations obey the criteria: 
\begin{align}
  \mathcal{S}_{in}=\left\{ \mathit{md}, \mathit{m=1,2}, \dots \mathit{M}_{1} \right\} \nonumber\\ 
       \mathcal{S}_{ou}=\left\{ \mathit{n}\left(\mathit{M+1}\right)\mathit{d}, \mathit{n=1,2}, \dots \mathit{M}_{2} \right\} 
\end{align}
For $\mathit{M}_{1}=\mathit{M}_{2}=\frac{\mathit{M}}{2}$, the resulting number of  physical sensors is even and yields $\frac{\mathit{M}^{2}-2}{2}+\mathit{M}$ DoFs.

Definition 2.4. A coprime pair of arrays \cite{Pal_2} is a sparse array with the sensors of two ULAs. The first one containing $\mathit{F}$ sensors with intersensor spacing $\mathit{Qd}$ and the other with $\mathit{2Q-1}$ sensors and intersensor gaps $\mathit{Fd}$.  The locations of their $\mathit{F +2Q-1}$ sensors are driven by 
\begin{align}
    \mathcal{S}=\left\{ \mathit{Qfd},\: 0\leq \mathit{f}\leq \mathit{F-1}  \right\} \cup \left\{\mathit{Fqd},\; 1\leq \mathit{q}\leq \mathit{2Q-1} \right\} 
\end{align}
A coprime pair of arrays provides up to $\mathit{2QF+1}$ DoFs \cite{Pal_2} using solely $\mathit{F+2Q-1}$ physical sensors. \vspace{-1em}
\subsection{System Model}
\label{Sub_sec_sys_model}
\begin{figure}[htb] 
	\centering 
	\includegraphics[width=7.6cm, height=5.0cm]{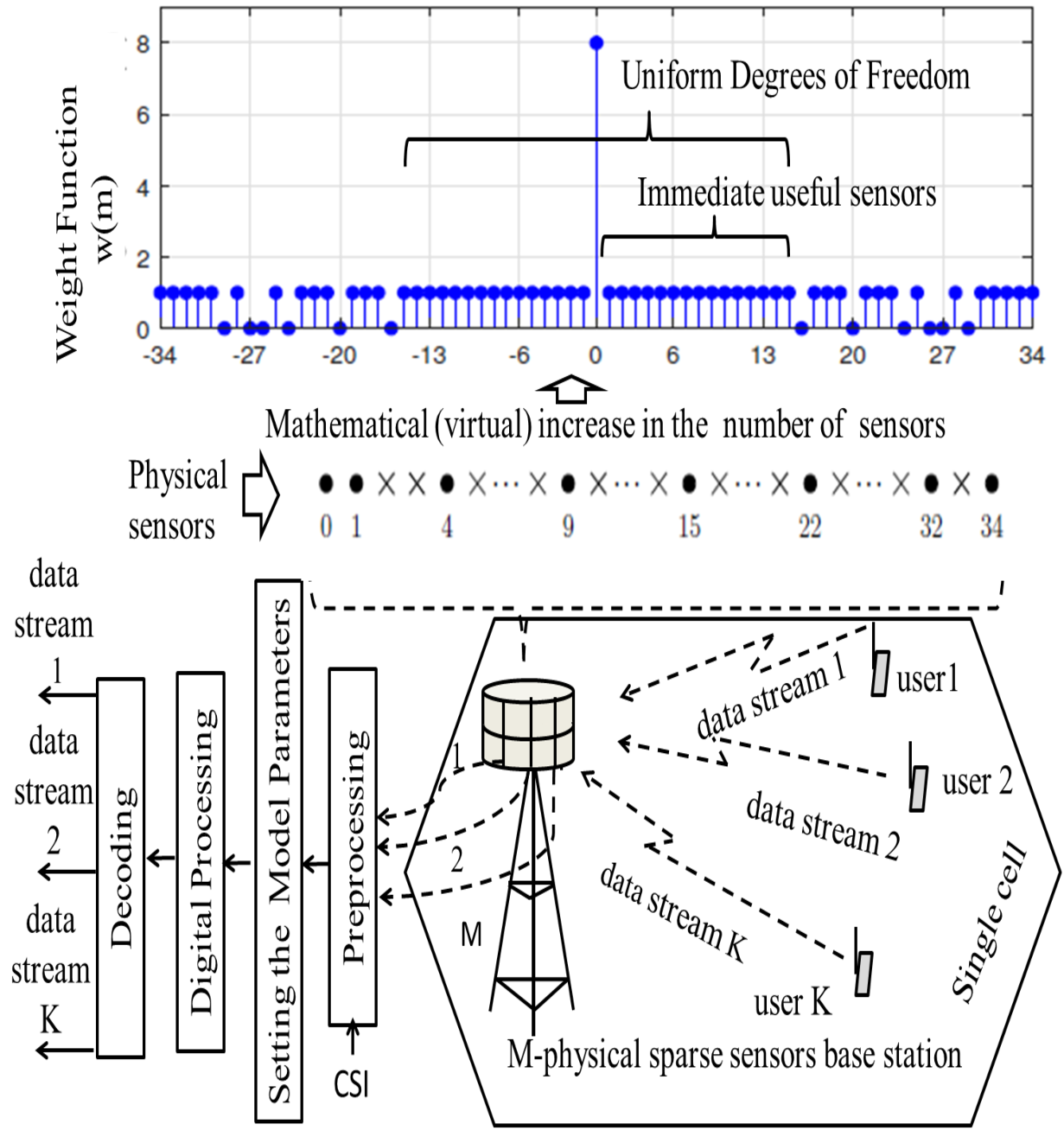} 
	\vspace{-0em}
	\caption{Uplink of a single-cell massive MIMO using sparse arrays}
	\label{Virtualization}
\end{figure}

Let us take into account the uplink of a single-cell MIMO system with a BS equipped with a sparse array composed by \textit{M} physical sensors, possibly a TLNA or CPA, on which the data symbols
$\lbrace \mathit{s}_{k} \rbrace$ $\in$ 
$\lbrace \mathit{s}_{1}, \mathit{s}_{2},\dots, \mathit{s}_{i=K} \rbrace$ included in radio frequency (RF) signals from $K$ single-antenna users, impinge on, as depicted in Fig.\ref{Virtualization}. The environment is assumed to be free of scatterers and, in contrast to massive MIMO settings, the number of physical sensors $M$ is assumed to be not much greater than the number of received users $K$. It is also assumed that all users transmit their data at the same time.

The received data vector $\mathbf{x}\left(t \right)$ $\in$ $\mathcal{C}^{M \times 1}$ at the BS is expressed by
\begin{align}
 \mathbf{x}\left(t \right)=  \mathbf{H} \mathbf{s}\left(t\right) + \mathbf{z}\left(t\right), \:\mathit{t=1,2,\cdots T\:\mathrm{snapshots} }
 \label{rec_data_vec}
\end{align}
where
 $\mathbf{x}\left(t \right)$ $\in$ $\mathcal{C}^{M \times 1}$ denotes the discrete-time received data vector and $\mathbf{z}\left(t\right)$ is the noise vector drawn from $\mathcal{CN}\left(\mathbf{0}, \sigma_{n}^{2}\mathbf{I}_{M}\right)$ and uncorrelated from the signal. The discrete-time signal vector
  $\mathbf{s}\left(t\right)$ $\in$ $\mathcal{C}^{K \times 1}$ contains the transmitted symbols. 
  The channel matrix $\mathbf{H}=\left[\mathbf{h}_{1},\mathbf{h}_{2},\dots, \mathbf{h}_{K}  \right]$  $\in$ $\mathcal{C}^{M \times K}$ is estimated at the receiver. 
  In \cite{Hawej,Hawej_1} a finite multipath channel model is devised for poor scattering environments, where the number of multipaths is much less than 
the number of BS antennas and users. In those works the channel vector  $\mathbf{h}_{k}\in\mathcal{C}^{\mathit{M\times1}} $ between the $k$th user and the BS is represented by
\vspace{-0.5em}
 \begin{align}
  \mathbf{h}_{k}= \frac{\beta_{k}}{\sqrt{\mathit{P}}}\sum_{p=1}^{P}\mathbf{a}\left(\theta_{\mathit{p}}\right)\mathit{g}_  {\mathit{kp}} 
  \label{propag_coef}
 \end{align}
where $\mathit{g}_{\mathit{kp}}$ is the random propagation gain coefficient between the $k$th user and the BS related to each path $\mathit{p}=\mathit{p}_{1}, \mathit{p}_{2},\dots, \mathit{P}$. In that studies $\mathit{g}_  {\mathit{kp}}$ is modeled as a Rayleigh fading coefficient with zero mean and unit variance, i.e., $\mathit{g}_{\mathit{kp}}$ $\sim$ $\mathcal{NC}\left(\mathrm{0}, 1\right)$. Furthermore, the expression \eqref{propag_coef} is simplified by the assumption that the path loss coefficient $\beta_{k}$ between the $k$th user and BS for all users is the same and normalized to unity. The steering vector  $\mathbf{a}\left(\theta_{\mathit{p}}\right)$ characterizes the angle measured between the direction of motion of the plane wave front and a line drawn perpendicular to the array. The variable $\theta_{\mathit{p}}$ $\in \left[-\frac{\pi}{2},\frac{\pi}{2} \right]$ denotes the angle of arrival corresponding to the \textit{p}th path, \textit{d} stands for the sensors spacing and $\lambda $ is the wavelength. However, that model assumes a uniform linear array (ULA) at BS and poor scattering, which results in possible multipaths.

Differently from the mentioned studies, the proposed  model assumes that there is no scattering, which means that there is only one path between the  $k$th user and the array. Thus, we can tailor \eqref{propag_coef} to the proposed model so that each path is associated to its user and its respective steering vector, similarly to line-of-sight propagation. Following this reasoning, it is possible to express the channel vector $\mathbf{h}_{k}$ as follows:
\begin{align}
  \mathbf{h}_{k}=\mathit{g}_{\mathit{k}}\:  \mathbf{a}\left(\theta_{\mathit{k}}\right)
  \label{propag_coef_mod}
 \end{align}
where $\mathit{g}_{\mathit{k}}$ $\sim$ $\mathcal{NC}\left(\mathrm{0}, 1\right)$ and $\mathbf{a}\left(\theta_{\mathit{k}}\right)$ is the steering vector that describes the Direction of Arrival (DOA) for each specific sparse array and user. 

It is known that sparse arrays can resolve $\mathcal{O}\left(\mathit{M}^{2}\right)$ signals as a result of the expanded difference coarrays obtained from differences of the distances between their sensors and the incidence of those differences. Nevertheless, some of the known  subtypes of this class of arrays like MRA and MHA neither present closed forms nor even have a simple formation rule for the design of the set containing the sensors' positions $\mathcal{S}$. Despite these drawbacks, 
it is possible to devise a method to be applied to TLNA and CPA, so that we can assess the theoretical performance of their class of arrays when applied to multiple-antenna communications. \vspace{-1em}
\section{Proposed Super-Resolution Multiple-Antenna Processing}
 We start the exposition of the proposed super-resolution multiple-antenna processing (PSRMAP) method by taking the expectation of the outer product of the received signal vector \eqref{rec_data_vec}, to obtain its covariance matrix $\mathbf{R}_{x}$ $\in \mathcal{C}^{M\times M}$ :   
\begin{align}
\mathbf{R}_{x}&=\mathcal{E}\left\{ \mathbf{x}\left(t\right)\: \mathbf{x}^{H}\left(t\right) \right\}= \sum_{k=1}^{K}\sigma_{k}^{2}\:\left(\mathbf{h}\left(\theta_{k}\right)\:\mathbf{h}^{H}\left(\theta_{k}\right)\right) + \sigma_{n}^{2}\mathbf{I}
\label{cov_R_x}
\end{align}
from which we obtain a vector with increased dimension 
\begin{align}
 \mathbf{v}= \mathrm{vec} \left(\mathbf{R}_{x}\right)= \left( \mathbf{H}^{*}\odot \mathbf{H}\right)\mathbf{p} + \sigma_{n}^{2} \vec{\mathbf{1}}_{n}, ~\in \mathcal{C}^{M^2}
 \label{vectorization}
\end{align}
where $\odot$ denotes the Khatri-Rao product, $\mathbf{p}=\left[\sigma_{1}^{2}, \sigma_{2}^{2},\dots,\sigma_{K}^{2}\right]^{T}$ and $\vec{\mathbf{1}}_{n}=\left[\mathbf{e}_{1}^{T}, \mathbf{e}_{2}^{T},\dots,\mathbf{e}_{M}^{ T}\right]^{T}$ 
stands for a column vector of all zeros except a $\mathit{1}$ in the $i$th position.
Both TLNA and CPA present  $\mathit{DoF}$ that allow the formulation of convenient non-negative definite matrices (NND), which can be combined with subspace methods to estimate a number of signals greater than the number of their respective receive physical sensors. In particular, a method like Multiple Signal Classification (MUSIC) combined with  a TLNA with an even number of physical sensors  can estimate up to $\frac{\mathit{M}^{2}}{4}+\frac{\mathit{M}}{2}-1$, signals. If TLNA is replaced by CPA, the estimation  can reach $\mathit{FQ}$ signals.
 
The vector $\mathbf{v}$ in \eqref{vectorization} obtained by vectorization of the  covariance matrix of the received signal $\mathbf{R}_{x}$ contains redundant information in the form of some elements that appear more than once. We can remove the repeated entries after their first appearance and arrange them so that the $i$th row coincides with the sensor situated at $\left(\frac{\mathit{-M}^{2}}{4}+\frac{\mathit{-M}}{2}+i\right)d$  and $\left(\mathit{-2FQ-1 +i} \right)d$ for TLNA and CPA, respectively.

The reshaped vectors obtained after the removal of these redundancies can be expressed for TLNA and CPA respectively as follows:
\begin{align}
 \mathbf{v}_{t,c}= \mathbf{B}_{t,c}\: \mathbf{p} +\mathbf{e}_{t,c}
 \label{reshap_long_vec_tlna}
\end{align}
%
where the augmented array manifolds  $\mathbf{B}_{t}=\left[\mathbf{q}_{t}\left(\theta_{1}\right), \mathbf{q}_{t}\left(\theta_{2}\right),\dots,\mathbf{q}_{t}\left(\theta_{K}\right)\right]$ and $\mathbf{B}_{c}=\left[\mathbf{q}_{c}\left(\theta_{1}\right), \mathbf{q}_{c}\left(\theta_{2}\right),\dots,\mathbf{q}_{c}\left(\theta_{K}\right)\right]$ are described by their steering vectors for TLNA and CPA, respectively, as
\begin{align}
\mathbf{q}_{t}\left(\theta_{k}\right)=&\lvert\mathit{g}_{\mathit{k}}\rvert^{2}\left[e^{-j2\pi\frac{d}{\lambda_{c}}\left(-\bar{M}+1 \right) \sin\theta_{k}}, e^{-j2\pi\frac{d}{\lambda_{c}}\left(-\bar{M}+2 \right) \sin\theta_{k}},\right.\nonumber\\
&\left.\ldots, e^{-j2\pi\frac{d}{\lambda_{c}}\left(\bar{M}-2 \right) \sin\theta_{k}}, e^{-j2\pi\frac{d}{\lambda_{c}}\left(\bar{M}-1 \right) \sin\theta_{k}} \right]^{T},
\label{augmented_steer_vec_tlna}
\end{align}
\begin{align}
\mathbf{q}_{c}\left(\theta_{k}\right)=&\lvert\mathit{g}_{\mathit{k}}\rvert^{2}\left[e^{-j2\pi\frac{d}{\lambda_{c}}\left(-\hat{M}+1 \right) \sin\theta_{k}}, e^{-j2\pi\frac{d}{\lambda_{c}}\left(-\hat{M}+2 \right) \sin\theta_{k}},\right.\nonumber\\
&\left.\ldots, e^{-j2\pi\frac{d}{\lambda_{c}}\left(\hat{M}-2 \right) \sin\theta_{k}}, e^{-j2\pi\frac{d}{\lambda_{c}}\left(\hat{M}-1 \right) \sin\theta_{k}} \right]^{T},
\label{augmented_steer_vec_cpa}
\end{align}
where $\mathit{k}=1,2, \dots, \mathit{K}$,  $\bar{M}=\frac{\mathit{M}^{2}}{4}+\frac{\mathit{M}}{2}$ and
$\hat{\mathit{M}}= \mathit{QF}+1$.
 The vectors $\mathbf{e}_{t}\in\mathcal{R}^{\left( 2\bar{M}-1\right)\times 1 }$  and $\mathbf{e}_{c}\in\mathcal{R}^{\left( 2\hat{M}-1\right)\times 1} $ consist of  all zeros, except for a $\mathit{1}$ in the central position.
In comparison with $\mathbf{x}$ in \eqref{rec_data_vec}, both $\mathbf{v}_{t}$ and $\mathbf{v}_{c}$ in \eqref{reshap_long_vec_tlna}   work as if the signals received  by a longer difference coarray whose sensors positions are computed by the diverse values in the set $ \left\{\vec{\mathit{x}}_{i}- \vec{\mathit{x}}_{j} \mid \mathit{i}\geq 1, \mathit{j}\leq M \right\}$, where $\vec{\mathit{x}}_{i}$ stands for the position vector of the  $i$th sensor. The source signal vector $\mathbf{p}$ composed by the powers $\sigma_{k}^{2}$, $\mathit{k}=1,2,\dots, K$ of the truly existing sources acts like coherent sources. This associated with the fact that the difference coarrays  are ULAs allow the application of spatial smoothing  to equations \eqref{reshap_long_vec_tlna}  to estimate full-rank covariance matrices, which are also NND. The  resulting  smoothed matrices obtained for TLNA \cite{Pal_1,Pinto_3} and for CPA \cite{Pal_2,Tan} can be expressed  respectively as follows:
\begin{align}
\bar{\mathbf{R}}_{tss}=\frac{1}{\bar{\mathit{M}}}\sum_{i=1}^{\bar{\mathit{M}}}\mathbf{v}_{t_i}\:\mathbf{v}^{H}_{t_i}=
\frac{1}{\bar{\mathit{M}}}\left[\left(\mathbf{B}_{t1}\:\Omega_{t}\:\mathbf{B}^{H}_{t1}\right) + \sigma_{n}^{2}\mathbf{I}_{\bar{\mathit{M}}}\right]^2 \label{cov_smoot_tlna}  
\end{align}
\vspace{-1.5em}
\begin{align}
\hat{\mathbf{R}}_{css}=\frac{1}{\hat{\mathit{M}}}\sum_{i=1}^{\hat{\mathit{M}}}\mathbf{v}_{c_i}\:\mathbf{v}^{H}_{c_i}=
\frac{1}{\hat{\mathit{M}}}\left[\left(\mathbf{B}_{c1}\:\Omega_{c}\:\mathbf{B}^{H}_{c1}\right) + \sigma_{n}^{2}\mathbf{I}_{\hat{\mathit{M}}}\right]^2
\label{cov_smoot_cpa}  
\end{align}
where the vector $\mathbf{v}_{t_i}$ is formed by the entries between the $\left(\bar{\mathit{M}}+1-i\right)$ and the $\left(2\bar{\mathit{M}}-i\right)$ rows of  $\mathbf{v}_{t}$ in \eqref{reshap_long_vec_tlna} and  $\mathbf{B}_{t1}$ denotes an array manifold composed with the final $\bar{M}$ rows of $\mathbf{B}_{t}$ in \eqref{reshap_long_vec_tlna}. Similarly,  $\mathbf{v}_{c_i}$ is composed with the entries included between the $\left(2\hat{\mathit{M}}+1-i\right)$ and the $\left(2\hat{\mathit{M}}-i\right)$ rows of  $\mathbf{v}_{t}$ in \eqref{reshap_long_vec_tlna} and  $\mathbf{B}_{c1}$ stands for an array manifold composed by the latest $\bar{M}$ rows of $\mathbf{B}_{c}$ in \eqref{reshap_long_vec_tlna}. 
More specifically,  covariance matrices $\Omega_{t}$  in \eqref{cov_smoot_tlna} and  $\Omega_{c}$ in \eqref{cov_smoot_cpa} respectively are expressed as $\Omega_{t}=\Omega_{c}= diag \left( \mathbf{p}\right)$.
Additionally, covariance matrices $\mathbf{B}_{t1}^{H}$  and  $\mathbf{B}_{c1}^{H}$ are described by
\begin{align}
\label{array_manifold_tlna_cpa}
 \mathbf{B}_{t1}^H,\mathbf{B}_{c1}^{H}= 
 \begin{bmatrix}
    1 & \mathit{\eta}_{1} & \cdots & \mathit{\eta}_{1}^{\left(\mathit{\gamma}\right)}\\
    1 & \mathit{\eta}_{2} & \cdots & \mathit{\eta}_{2}^{\left(\mathit{\gamma}\right)}\\
    \vdots & & \ddots & \\
    1 & \mathit{\eta}_{K} & \cdots & \mathit{\eta}_{K}^{\left(\mathit{\gamma}\right)} 
 \end{bmatrix}
\end{align}
where $\mathit{\gamma}= \left(\bar{\mathit{M}}-1 \right)$ for $\mathbf{B}_{t1}^H$ and 
$\mathit{\gamma}= \left(\hat{\mathit{M}}-1\right) $ for $\mathbf{B}_{c1}^H$.
The matrices $\bar{\mathbf{R}}_{t}=\left(\bar{\mathbf{R}}_{tss}\right)^{1/2}$, for TLNA, and $\bar{\mathbf{R}}_{c}=\left(\bar{\mathbf{R}}_{css}\right)^{1/2}$, for CPA, play a pivotal role in this work. Their augmented sizes, which are results of preprocessed signals collected by $M$ and $\mathit{F +2Q}-1$ sparse physical sensors, provide resolution of up to $\frac{\mathit{M}^{2}}{4}+\frac{\mathit{M}}{2}-1$ and $\mathit{QF}$ signals, respectively.
\section{Proposed MMSE Receivers}
We can use the enlarged matrices $\bar{\mathbf{R}}_{t}$ and  $\bar{\mathbf{R}}_{c}$ defined previously to determine the expressions of the augmented data vectors that give origin to them. %
Now,  let us assume that our system model  is similar to that described in Subsection \ref{Sub_sec_sys_model}, except that our array is a ULA, composed with $\mathit{J}\lvert_{\mathit{=\bar{M} ;\hat{M}}}$ in \eqref{augmented_steer_vec_tlna} and \eqref{augmented_steer_vec_cpa} sensors, which is the same number of equivalent sensors obtained at the end of the sparse (TLNA or CPA) array preprocessing. It receives signals transmitted {simultaneously} from the $\mathit{K}$ single users  assumed  in Subsection \ref{Sub_sec_sys_model}.
Following this interpretation, we have  that the received data vector $\mathbf{x}_{U}\left(t \right)$ $\in$ $\mathcal{C}^{J \times 1}$ at the BS would be expressed by
\begin{align}
 \mathbf{x}_{U}\left(t \right)=  \mathbf{H}_{U}\: \mathbf{s}\left(t\right) + \mathbf{z}\left(t\right), \:\mathit{t=1,2,\cdots T\:\mathrm{snapshots} }
 \label{rec_data_vec_2}
\end{align}
where $\mathbf{H}_{U}=\left[\mathbf{h}_{1},\mathbf{h}_{2},\dots, \mathbf{h}_{K}  \right]$  $\in$ $\mathcal{C}^{J \times K}$ stands for the channel,  $\mathbf{z}\left(t\right)$  denotes the noise vector, assumed to be $\mathcal{CN}\left(\mathbf{0}, \sigma_{n}^{2}\mathbf{I}_{J}\right)$. It is assumed to be uncorrelated from the signal. 
This long ULA whose number of  physical sensors  is equal to the number of the equivalent  virtual sensors  obtained by preprocessing CPA and TLNA  presents the following received data covariance matrix: 
\begin{align}
\mathbf{R}_{U}&=\mathcal{E}\left\{\mathbf{x}_{U}\left(t\right)\:\: \mathbf{x}_{U}^{H}\left(t\right)\right\}= \mathbf{H}_{U}\:\mathbf{R}_{ss}\:\mathbf{H}_{U}^{H}+ \sigma_{n}^{2} \mathbf{I}_{J}
\label{cov_eq_ULA_for_comp}
 \end{align}
Now, we can rewrite $\bar{\mathbf{R}}_{t}$ for TLNA, and $\bar{\mathbf{R}}_{c}$, for CPA, in compact form, as follows: 
\begin{align}
 \bar{\mathbf{R}}_{t,c}=\left(\bar{\mathbf{R}}_{tss,css}\right)^{1/2} &=  
\frac{1}{{\mathit{\sqrt{J}}}}\left[\left(\mathbf{B}_{t1,c1}\:\Omega_{t,c}\:\mathbf{B}^{H}_{t1,c1}\right) + \sigma_{n}^{2}\mathbf{I}_{\mathit{J}}\right] \nonumber\\&= \mathcal{E}\left\{\mathbf{x}_{a}\left(t\right)\quad \mathbf{x}_{a}^{H}\left(t\right)\right\}
\label{cov_eq_tlna_and_cpa_for_comp}
\end{align}
where
\begin{align}
\mathbf{x}_{a}\left(t\right) & = \mathit{J}^{-\frac{1}{4}} \left(\mathbf{B}_{t1,c1}\:\mathbf{s}_{a}\left(t\right) + \mathbf{z}_{a}\left(t \right) \right)
\label{data_vec_long_equiv_ULA}
\end{align}
By comparing \eqref{cov_eq_tlna_and_cpa_for_comp} and \eqref{cov_eq_ULA_for_comp}, it can be noticed that sparse augmented covariance matrices and similar corresponding to equivalent elongated ULAs possess the same number of equally spaced sensors, and for this reason, result in equivalent channel vectors. They also present  transmit uncorrelated signals with different amplitudes but have the same SNR. As a matter of fact, except for the factor $\frac{1}{{\mathit{\sqrt{J}}}}$ in the enlarged covariance expressions of the sparse arrays \eqref{cov_eq_tlna_and_cpa_for_comp}, expressions \eqref{cov_eq_ULA_for_comp} and \eqref{cov_eq_tlna_and_cpa_for_comp}  are similar. 
Now, let us consider the  
estimate  of a data symbol $\mathbf{s}_{a}$ of the augmented discrete-time received data vector $\mathbf{x}_{a}(t)$ \eqref{rec_data_vec} using a suitable
receive filter $\mathbf{w}_{a}\left(t\right)$ whose output is expressed by
\begin{align}
\hat{\mathit{s}}_{a}\left(t\right)=\mathbf{w}_{a}^H\left(t\right)\:\mathbf{x}_{a}\left(t\right),
\label{beamformer_definition}
\end{align}
where $1\leq\mathit{t}\leq \mathit{T}$ snapshots. 
Equation \eqref{data_vec_long_equiv_ULA}, can be rewritten in terms of the desired signal and interferences, as follows:
\begin{align}
 \mathbf{x}_{a}\left(t\right)&=  \mathit{J}^{-\frac{1}{4}}\left( \mathbf{b}_{k} \mathit{s}_{k}\left(t\right)+\sum_{\substack{i=1;\:i\neq\mathit{k}}}^{K}\mathbf{b}_{i} \mathit{s}_{i}\left(t\right) + \mathbf{z}_{J}\left(t\right)\right) \nonumber\\ &=\mathit{J}^{-\frac{1}{4}}\left(\mathbf{b}_{k} \mathit{s}_{k}\left(t\right)+\mathbf{i}\left(t\right) +\mathbf{z}_{J}\left(t\right)\right)
 \label{signal_expansion}
\end{align}
where the vectors $\mathbf{b}_{k}$, $\mathbf{b}_{i}$ and $\mathbf{z}_{J}$ $\in$ $\mathcal{C}^{\mathit{J}\times 1} $ denote the channel vectors corresponding to the desired user, the interferences, both obtained from the augmented channel matrices in \eqref{array_manifold_tlna_cpa} and the noise vector corresponding to TLNA or CPA. It can be noticed that $\mathbf{i}\left(t\right)$ is equal to the  summation of the interferences combined with their respective channel vectors, i.e., $\sum_{\substack{i=1\\i\neq\mathit{k}}}^{K}\mathbf{b}_{i} \mathit{s}_{i}\left(t\right)$. 
It can be shown that the Mean-Squared Error (MSE) between the transmit symbol and its estimated value  \cite{Stoica_Li}  can be expressed by the following expectation:
\begin{align}
 \mathcal{E}\left\{\lvert \hat{\mathit{s}}-\mathit{s}\rvert^{2} \right\}&=
 \mathit{J}^{-\frac{1}{2}}\sigma_{k}^{2}\mathbf{w}^{H}_a\mathbf{b}_{k}\mathbf{b}_{k}^{H}\mathbf{w}_a - \mathit{J}^{-\frac{1}{4}}\:\sigma_{k}^{2}\mathbf{w}^{H}_a \mathbf{b}_{k} \nonumber\\& - \mathit{J}^{-\frac{1}{4}}\:\sigma_{k}^{2}\mathbf{b}_{k}^{H}\mathbf{w}_a +\sigma_{k}^{2}+ \mathit{J}^{-\frac{1}{2}}\mathbf{w}^{H}_a \mathbf{R}_{\mathit{i +n}}\mathbf{w}_a
 \label{MSE_final}
\end{align}
where
\begin{align}
\mathbf{R}_{\mathit{i +n}}&= \mathcal{E}\left\{\mathbf{i}\:\mathbf{i}^{H}\right\}+ \:\mathcal{E}\left\{\mathbf{i}\:\mathbf{z}^{H}\right\}\rvert_{=0} +\mathcal{E}\left\{\mathbf{z}\:\mathbf{i}^{H}\right\}\rvert_{=0}+  \:\mathcal{E}\left\{\mathbf{z}\:\mathbf{z}^{H}\right\}
\label{cross_terms_MMSE}
\end{align}
is the equivalent interference-plus-noise covariance matrix, where the cross terms indicated in \eqref{cross_terms_MMSE} are equal to  zero matrices, according to the assumed  uncorrelation between signal and noise.
By differentiating the MSE in \eqref{MSE_final} with respect to the receive filter $\mathbf{w}$ and making it  equal to $0$, we can compute the minimum MSE (MMSE) receive filter, as follows: 
\begin{align}
   \mathit{J}^{-\frac{1}{2}}\sigma_{k}^{2}\:\left(\mathbf{b}_{k}\:\mathbf{b}_{k}^{H}\right)^{T}\mathbf{w}_a^{*} - \mathit{J}^{-\frac{1}{4}}\sigma_{k}^{2}\: \mathit{b}_{k}^{*}  + \mathit{J}^{-\frac{1}{2}}\mathbf{R}_{\mathit{i +n}}^{T}\:\mathbf{w}_a^{*}=0
  \label{MSE_derivative}  
\end{align}
%
Note that the linear MMSE receive filters can be employed by many detectors employed at the BS for uplink receive processing techniques \cite{jidf,spa,stmf,mbsic,wlmwf,mfsic,dfcc,jiomimo,jiols,tds_cl,smtvb,tds,mbdf,bfidd,1bitidd,armo,did,sicdma,vfap,rrber,memd,baplnc,rsrbd,dynovs,aaidd,iddmtc,cpm,mtcdet,1bitce,msgamp1,msgamp2,comp}.

\subsection{Linear MMSE Receivers}
In this subsection, we describe linear MMSE receivers with super-resolution, whose column vectors of its matrix form  $\mathbf{W}_{a}=\left[\mathbf{w}_{a_1}, \mathbf{w}_{a_2},\dots, \mathbf{w}_{a_K}\right]$ can be obtained by solving \eqref{MSE_derivative}: 
\begin{align}
\mathbf{w}_{a_i}=\mathit{J}^{\frac{1}{4}}\:\sigma_{k}^{2}\left(\mathbf{R}_{\mathit{i +n}}+\: \sigma_{k}^{2}\:\mathbf{b}_{k}\mathbf{b}_{k}^{H}\right)^{-1}\mathbf{b}_{k} ~\in \mathcal{C}^{J\times1}
\label{desired_weight_vector_1}
\end{align}
Assuming  perfect channel state information and that the statistical knowledge of the noise is available, the problem corresponds to estimating the transmitted symbol and then performing detection, which can be extracted from the estimate of the data symbol  $\tilde{\mathbf{s}}_{a}\left(\mathit{t}\right)$, which in turn can be expressed as follows:
\begin{align}
 \tilde{\mathit{s}}_{a}\left(\mathit{t}\right)= {\rm Slicer} \left[ \hat{\mathit{s}}_{a}\left(\mathit{t}\right)\right] 
\end{align} 
where $\hat{\mathit{s}}_{a}\left(\mathit{t}\right)$ is computed by \eqref{beamformer_definition}.
\subsection{Successive Interference Cancellation MMSE Receivers}
\label{SIC_MMSE_receivers}
The interference cancellation can be successfully combined with an MMSE filter to better evaluate the  performance of sparse arrays in terms of uncoded BER. Specifically, in the norm-based ordered successive interference cancellation (OSIC) method \cite{Paulraj, Wolniansky}, which will be applied to our PSRMAP it is assumed that the received signal concentration of the \textit{i}th transmitted signal is
proportional to the norm of its corresponding channel vector $\left[\mathbf{B}_{t1}\right]_{\left(:,i\right)}$. Thus, before starting the procedure, the channel vectors are norm-ordered decreasingly.  
Except for the first step, which preserves the MMSE linear filter features, the subsequent refining steps  remove the preceding interference according the decreasing order of the Euclidean norm of that vectors. 
The second step of OSIC is provided by the $\mathit{2}$nd row of the MMSE linear filter applied to the first 'peeled' received vector $\tilde{\mathbf{x}}_{1}$, which is the difference between the received vector $\mathbf{x}_{a}$  and  the product of the first norm-ordered
channel vector $\mathit{J}^{-\frac{1}{4}}\mathbf{b}_{t1,c1\left(1\right)} $ and its corresponding signal estimate $\mathit{s}_{a_1}\left(t\right)$. The process for estimating the 'peeled' received vector starts with  rewriting \eqref{data_vec_long_equiv_ULA} as follows %
\begin{align}
 \mathbf{x}_{a}\left(t\right) & = \mathit{J}^{-\frac{1}{4}}\left(\mathbf{b}_{t1,c1\left(1\right)}  \:\mathit{s}_{a_1}\left(t\right) +\mathbf{b}_{t1,c1\left(2\right)}  \:\mathit{s}_{a_2}\left(t\right) \right.\nonumber\\
&\left.+ \ldots+\mathbf{b}_{t1,c1\left(K\right)}  \:\mathit{s}_{a_K}\left(t\right) + \mathbf{z}_{a}\left(t \right) \right) 
\end{align}
and after that, taking the desired difference:
\begin{align}
 \tilde{\mathbf{x}}_{1}&=\mathbf{x}_{a}\left(t\right)- \mathit{J}^{-\frac{1}{4}} \mathbf{b}_{t1,c1\left(1\right)}  \:\mathit{s}_{a_1}\left(t\right)\nonumber\\ &= \mathit{J}^{-\frac{1}{4}}\left(\mathbf{b}_{t1,c1\left(2\right)}  \:\mathit{s}_{a_2}\left(t\right) \right.\nonumber\\
&\left.+ \ldots+\mathbf{b}_{t1,c1\left(K\right)}  \:\mathit{s}_{a_K}\left(t\right) + \mathbf{z}_{a}\left(t \right) \right)
\end{align}
%
%
\section{Analysis}
\vspace{-0.5em}

In this section, we analyze aspects of the performance of the proposed super-resolution multiple-antenna processing. To this end,  we assume Gaussian signalling and make use of 
suitable indicators such as Achievable Sum-Rate $\left(\mathit{ASR}_{a}\right)$. 
The  computational complexity $\left( \mathit{CC}_{a}\right)$ demanded for carrying out the proposed  algorithm follows $\mathcal{O}\left(\mathit{J}^{3}\right)$ Floating Point Operations per second (FLOPs) mainly concentrated in its inversion matrix. 
%
%
The $\mathit{ASR}$ can be defined \cite{Marzetta,Hawej} as:   
\begin{equation}
    \mathit{ASR}_{a}\leq \sum_{k=1}^{K} \log_{2}\left( 1 + \mathit{SINR}_{k}\right)
    \label{def_ASR}
\end{equation}
in which the received signal-to-interference plus noise ratio of the \textit{k}th user ($\rm SINR_{k}$), which is taken at the output of the receiver is expressed by:
 \begin{align}
  \mathit{SINR}_{k}= \dfrac{ \sigma_s^2}{\sigma_j^2 + \sigma_w^2}
  \label{def_sinr_usu}
 \end{align}
where $\sigma_s^2$ is the signal power, $\sigma_j^2$ is the interference power and $\sigma_w^2$ is the noise power. Recall that  as our system is composed of a sole cell, the inter-cell interference is not considered. 
This ratio can be obtained by identifying  and taking suitable parts of the expression of the variance of the  estimate of the data symbol $\gamma= \mathcal{E}\left\{ \hat{\mathit{s}}_{a}\left(t\right)\hat{\mathit{s}}^{H}_{a}\left(t\right)\right\}$ as follows:
\begin{align}
 \gamma &=\mathit{J}^{-\frac{1}{2}}
 \bigl(\overbrace{\sigma_{k}^{2}\lvert \mathbf{w}_{k}^{H}\: \mathbf{b}_{k}\rvert^{2}}^{\textit{signal power }} +\:\overbrace{\sum_{\substack{i=1;i\neq\mathit{k}}}^{K}\sigma_{i}^{2} \lvert \mathbf{w}_{k}^{H}\: \mathbf{b}_{i}\rvert^{2}}^{\substack{\textit{ intra-cell MUI }}}   +\: \overbrace{\sigma_{n}^{2} \norm {\mathbf{w}_{k}}^{2}}^{\textit{noise }} \bigr)
 \label{parts_of_ASR}
\end{align}
where  we have considered that the expected values of cross-product terms are null as a result of the statistics assumed in Subsection \ref{Sub_sec_sys_model}. From \eqref{def_ASR}, \eqref{def_sinr_usu}  and \eqref{parts_of_ASR}, we obtain the expression of $\mathit{ASR}_{a}$  for  the considered sparse arrays, namely, \textit{TLNA} and \textit{CPA}, as follows: %
\begin{align}
  \mathit{ASR}_{a}\leq \sum_{k=1}^{K} \log_{2}\Bigl( 1 + \frac{\sigma_{k}^{2}\lvert \mathbf{w}_{k}^{H}\: \mathbf{b}_{k}\rvert^{2}}{\sum_{\substack{i=1;\:i\neq\mathit{k}}}^{K}\sigma_{i}^{2} \lvert \mathbf{w}_{k}^{H}\: \mathbf{b}_{i}\rvert^{2} + \sigma_{n}^{2} \norm {\mathbf{w}_{k}}^{2}}\Bigr)
  \label{ASR_sparse_arrays}
\end{align}
%
%
%
%
\section{Numerical results}
\label{numerical_results}

This section is intended to assess the performance of our PSRMAP from the point of view of the ASR and BER. 
To this end, we examine a scenario involving  $\mathit{K=8}$ single-antenna users. The transmitted signals containing $\mathit{10}^{2}$ symbols under QPSK modulation experience a channel modeled as in \eqref{propag_coef_mod}  before impinging on TLNA, CPA and ULA-based MU-MIMO receivers. The first and the second arrays comprise $\mathit{M}\in \left\{ 8,16\right\} $ non-uniformly spaced physical sensors  whereas ULA  consists of $\mathit{M} =  16 $ uniformly spaced physical sensors. We set the number of independent trials to $\mathit{10}^{3}$.

\begin{figure}[htb!]
	\centering
	\begin{minipage}{0.525\columnwidth}
		\centering
		\includegraphics[width=\textwidth,height=3.6cm]{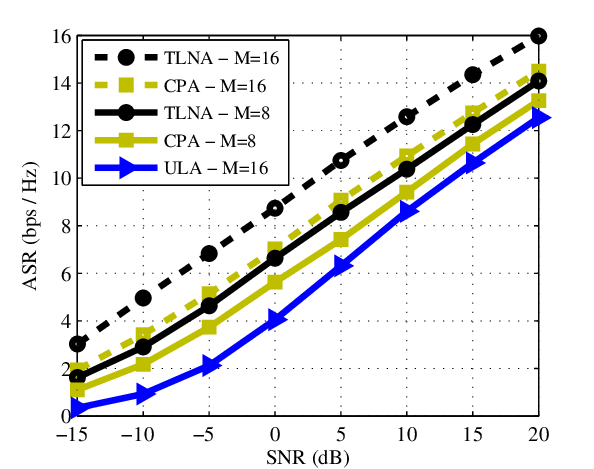}
		\caption{ASR for $\mathit{K}=8$ single\\ users. M= No. of physical sensors.}
		\label{ASR_nest_cop_8_rx_8_tx_ula_16_rx_8_tx_1000_runs}
	\end{minipage}%
	\begin{minipage}{0.525\columnwidth}
		\centering
		\includegraphics[width=\textwidth,height=3.6cm]{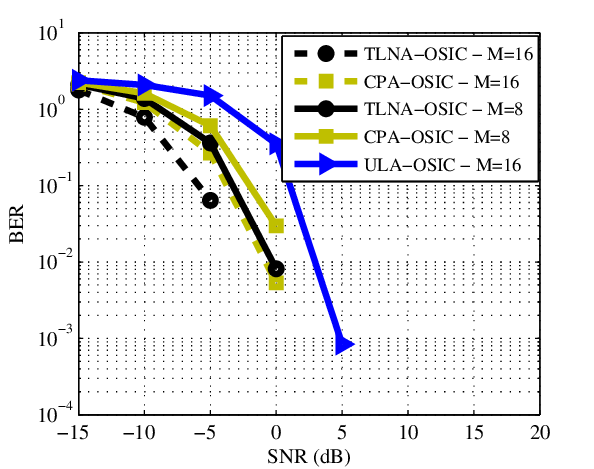}
		\caption{Uncoded BER for $\mathit{K}=8$ single\\ users. M= No. of physical sensors.  }
		\label{BER_nest_cop_8_rx_8_tx_ula_16_rx_8_tx_1000_runs_OSIC}
	\end{minipage}
\end{figure}

 In Fig.\ref{ASR_nest_cop_8_rx_8_tx_ula_16_rx_8_tx_1000_runs}, we plot the ASR corresponding to  $\mathit{K=8}$ users sending signals to a ULA comprising  $\mathit{M}=16$ physical sensors. We can compare its ASR to that achieved by TLNA and CPA consisting both of an smaller or equal amount of sensors, i.e.  $\mathit{M}=8$ and $\mathit{M}=16$. It can be noticed that the smaller number of sensors for TLNA and CPA $\left(\mathit{M}=8\right)$ is just enough to provide better performance than that achieved by a ULA containing $\left(\mathit{M}=16\right)$  in all considered SNR range.  %

Fig.\ref{BER_nest_cop_8_rx_8_tx_ula_16_rx_8_tx_1000_runs_OSIC} illustrates the performance of the uncoded BER  for TLNA, CPA and ULA  under  norm-ordered sucessive interference cancellation applied to a MMSE receiver, as described in Subsection \ref{SIC_MMSE_receivers}. A comparison among the curves makes clear that for TLNA and CPA with the same number of physical sensors, the first yields better results in the SNR range in question. Furthermore, the BER performance of a ULA consisting of $\mathit{M}=16$ is worse than that of a  pair $\left\{TLNA, CPA\right\}$ comprised of $\mathit{M}=8$, i.e, its  half, and also by the same pair consisting of the same number of sensors of a ULA $\mathit{M}=16$.   
\vspace{-1.00em}
\vspace{-2.00em}
\section{Conclusions}
We have proposed a sparse arrays processing  common to two-level nested and co-prime arrays that can be applied to multiple-antenna systems in the light of the properties of  the similar virtual ULAs. It is assumed  a geometry-based stochastic model and  no scattering inside a single-cell. The proposed PSRMAP procedure resulted in substantial gains in terms of achievable rates, bit error ratio and the energy savings resulting from the much smaller number of sensor elements which is required to achieve the same performance of a given ULA. The obtained results motivate further studies in this area. 

\end{document}